\documentclass[prb,twocolumn,superscriptaddress,preprintnumbers,amsmath,amssymb,showpacs]{revtex4}

\usepackage{graphicx}
\usepackage{amsmath}
\usepackage{latexsym}
\usepackage{dcolumn}
\usepackage{bm}
\usepackage{float}
\usepackage{graphicx,here}

\begin{document}

\title{Spin Texture in Type-\uppercase\expandafter{\romannumeral2} Weyl Semimetal WTe$_2$}

\author{Baojie Feng}
\thanks{bjfeng@issp.u-tokyo.ac.jp}
\affiliation{Institute for Solid State Physics, The University of Tokyo, Kashiwa, Chiba 277-8581, Japan}
\author{Yang-Hao Chan}
\affiliation{Institute of Atomic and Molecular Sciences, Academia Sinica, Taipei 10617, Taiwan}
\author{Ya Feng}
\affiliation{Institute of Physics, Chinese Academy of Sciences, Beijing 100190, China}
\author{Ro-Ya Liu}
\affiliation{Institute for Solid State Physics, The University of Tokyo, Kashiwa, Chiba 277-8581, Japan}
\author{Mei-Yin Chou}
\affiliation{Institute of Atomic and Molecular Sciences, Academia Sinica, Taipei 10617, Taiwan}
\affiliation{School of Physics, Georgia Institute of Technology, Atlanta, GA 30332, USA}
\affiliation{Department of Physics, National Taiwan University, Taipei 10617, Taiwan}
\author{Kenta Kuroda}
\affiliation{Institute for Solid State Physics, The University of Tokyo, Kashiwa, Chiba 277-8581, Japan}
\author{Koichiro Yaji}
\affiliation{Institute for Solid State Physics, The University of Tokyo, Kashiwa, Chiba 277-8581, Japan}
\author{Ayumi Harasawa}
\affiliation{Institute for Solid State Physics, The University of Tokyo, Kashiwa, Chiba 277-8581, Japan}
\author{Paolo Moras}
\affiliation{Istituto di Struttura della Materia, Consiglio Nazionale delle Ricerche, I-34149 Trieste, Italy}
\author{Alexei Barinov}
\affiliation{ELETTRA-Sincrotrone Trieste S.C.p.A., 34149 Basovizza, Trieste, Italy}
\author{Walid Malaeb}
\affiliation{Institute for Solid State Physics, The University of Tokyo, Kashiwa, Chiba 277-8581, Japan}
\affiliation{Physics Department, Faculty of Science, Beirut Arab University, P. O. Box 11-5020 Beirut, Lebanon}
\author{C$\rm\acute{e}$dric Bareille}
\affiliation{Institute for Solid State Physics, The University of Tokyo, Kashiwa, Chiba 277-8581, Japan}
\author{Takeshi Kondo}
\affiliation{Institute for Solid State Physics, The University of Tokyo, Kashiwa, Chiba 277-8581, Japan}
\author{Shik Shin}
\affiliation{Institute for Solid State Physics, The University of Tokyo, Kashiwa, Chiba 277-8581, Japan}
\author{Fumio Komori}
\affiliation{Institute for Solid State Physics, The University of Tokyo, Kashiwa, Chiba 277-8581, Japan}
\author{Tai-Chang Chiang}
\affiliation{Department of Physics, National Taiwan University, Taipei 10617, Taiwan}
\affiliation{Department of Physics, University of Illinois, Urbana, IL 61801, USA}
\author{Youguo Shi}
\affiliation{Institute of Physics, Chinese Academy of Sciences, Beijing 100190, China}
\author{Iwao Matsuda}
\thanks{imatsuda@issp.u-tokyo.ac.jp}
\affiliation{Institute for Solid State Physics, The University of Tokyo, Kashiwa, Chiba 277-8581, Japan}

\date{\today}


\begin{abstract}
We determine the band structure and spin texture of WTe$_2$ by spin- and angle-resolved photoemission spectroscopy (SARPES). With the support of first-principles calculations, we reveal the existence of spin polarization of both the Fermi arc surface states and bulk Fermi pockets. Our results support WTe$_2$ to be a type-\uppercase\expandafter{\romannumeral2} Weyl semimetal candidate and provide important information to understand its extremely large and non-saturating magnetoresistance.
\end{abstract}

\maketitle


Weyl semimetals, a novel state of topological quantum matter, have attracted significant attention in the recent years\cite{Hosur2012,WengH2015,XuSY2015TaAs,LvBQ2015,YangLX2015}. The low-energy electronic excitations (quasiparticles) in Weyl semimetal behave as Weyl fermions, a long-sought fundamental particle that has not been discovered until now. In a Weyl semimetal, the Weyl points always appear in pairs with opposite chirality, and can be described as magnetic monopoles in the momentum space. Near each Weyl point, the bands disperse linearly along all three momentum directions, thus forming three-dimensional Weyl cones. Besides the intriguing bulk bands, there exist topological non-trivial Fermi arcs that connect the projections of bulk Weyl points on the surface. The Weyl semimetals can be further classified into two types. Type-\uppercase\expandafter{\romannumeral1} has a point-like Fermi surface with symmetric Weyl cones, which has been realized in the TaAs family\cite{XuSY2015TaAs,LvBQ2015,YangLX2015,XuSY2015NbAs,LiuZK2016,LvBQ2015spin,XuSY2015spin}; in the type-\uppercase\expandafter{\romannumeral2}, the Lorentz invariance is strongly violated and the Weyl cones, tilted over one side, appear at the contact points between electron and hole pockets\cite{Soluyanov2015}. Recently, evidence of type-\uppercase\expandafter{\romannumeral2} Weyl semimetal has been reported in LaAlGe\cite{XuSY2016LAG}, MoTe$_2$\cite{JiangJ2016,LiangA2016,DengK2016,HuangL2016,XuN2016} and WTe$_2$\cite{Bruno2016,WangC2016,WuY2016}. However, the Weyl semimetal character of WTe$_2$ has not yet been validated because neither the bulk Weyl points nor the non-trivial Fermi arcs has been experimental confirmed.

\begin{figure}[!b]
\includegraphics[width=8.5cm]{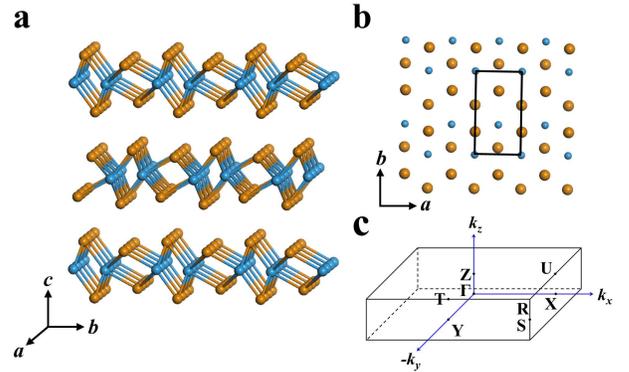}
\caption{(a) Three dimensional view of the crystal structure of WTe$_2$. The blue and orange balls represent W and Te atoms, respectively. (b) Top view, $\textit{i.e.}$, across the $\mathit{c}$ axis. The W-W zigzag chains are in the $\mathit{a}$ axis. The black rectangle indicates a surface unit cell. (c) Schematic drawing of the three-dimensional Brillouin zone (BZ) of WTe$_2$ with the high symmetry points indicated.}
\label{fig1}
\end{figure}

\begin{figure*}[!t]
\includegraphics[width=17.2 cm]{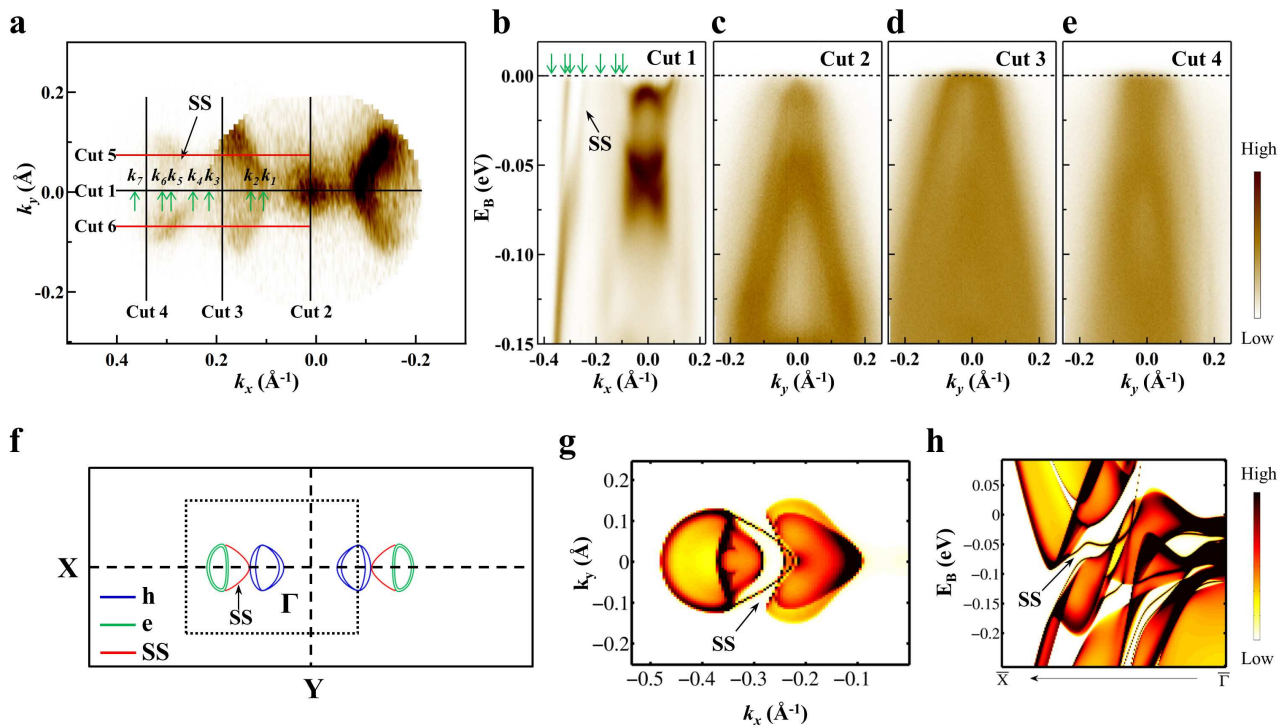}
\caption{(a) Fermi surface of WTe$_2$ integrated within 10 meV at the Fermi level. The label ``SS" mark the Fermi arc surface states. (b-e) ARPES intensity plots along cut 1 to cut 4 marked in (a). The green arrows in (a) and (b) indicate the bands that cross the Fermi level. (f) Schematic drawing of the Fermi surface. The blue, green and red lines indicate the hole pockets, electron pockets and Fermi arcs, respectively. The black dashed rectangle indicates the area measured in (a). (g) Momentum-resolved surface density of states at the Fermi level based on the contributions from the first unit cell (two triple layers). (h) Surface density of states as a function of energy along the $\Gamma$-X direction.}
\label{fig2}
\end{figure*}

Weyl semimetals are expected to host novel properties such as quantum anomalous Hall effect, extremely large magnetoresistance (XMR), and so forth\cite{Ashby2013,Hosur2012',Landsteiner2014,ZhangC2015,Shekhar2015}. Indeed, before being found as a type-\uppercase\expandafter{\romannumeral2} Weyl semimetal, WTe$_2$ has already been reported to host XMR which can reach 13 million percent in a magnetic field of 60 T without signature of saturation\cite{Ali2014}. The origin of XMR in WTe$_2$ is still unclear until now. One explanation is that electrons and holes perfectly compensate in WTe$_2$\cite{Ali2014,Lv2015,Pletikosic2014,Cai2015,Zhao2015,Xiang2015}, but later experiments by high-resolution angle resolved photoemission spectroscopy (ARPES) and magneto-transport measurements revealed that the electron and hole densities are slightly imbalanced\cite{Jiang2015,Rhodes2015,WangYL2016}. Recently, new mechanisms for XMR have been proposed based on the dynamics of spin-polarized electrons under the external magnetic field. For example, D. Rhodes \textit{et al.}\cite{Rhodes2015} suggest that the electronic structure of WTe$_2$ is composed of spin-split bands due to the spin-orbit interaction. The spin-polarized Fermi surface evolves sensitively with the magnetic field due to the Zeeman effect. On the other hand, J. Jiang \textit{et al.}\cite{Jiang2015} propose that the resistivity of WTe$_2$ is intrinsically small due to the prohibition of backscattering between spin-polarized Fermi pockets and the XMR effect is realized by the breakdown of this restriction due to the change of the spin texture in external magnetic field. The existence of spin-polarized bands in WTe$_2$ has recently been confirmed by SARPES\cite{Das2016}, but detailed studies of the spin texture, necessary to understand the mechanism of XMR, are still lacking.

In this Letter, we perform detailed SARPES to directly measure the band structure and spin texture of WTe$_2$. Combined with first-principles calculations, we unravel the existence of spin-polarized Fermi arcs that connect the opposite Weyl points. Moreover, we find that the Fermi pockets are also spin-polarized, in agreement with our theoretical calculations. Our results provide strong evidence for the type-\uppercase\expandafter{\romannumeral2} Weyl semimetal states in WTe$_2$ and give important information on the mechanism of XMR in this material.

Single crystals WTe$_2$ were grown by solid-state reaction methods and the details have been described elsewhere\cite{Kangd2015}. The samples were cleaved under ultra-high vacuum with a base pressure better than 1$\times$10$^{-10}$ mbar. SARPES measurements were performed at the Institute for Solid State Physics, the University of Tokyo\cite{YajiK2016}. The photoelectrons were excited by a laser source (h$\rm\nu$=6.994 eV) and measured by a hemispherical analyzer (ScientaOmicron DA30-L). Twin Very-Low-Energy-Electron-Diffraction (VLEED) detectors were equipped to determine the spin texture. All the photoemission data were taken using a \textit{p}-polarized light with a sample temperature of 7 K. In the spin-integrated ARPES mode, the energy and angle resolutions were 1.5 meV and 0.1$^{\circ}$, respectively; in the spin-resolved ARPES mode, the energy and angle resolutions were 20 meV and 0.7$^{\circ}$, respectively.

First-principles calculations was performed with the Vienna Ab initio Simulation package (VASP)\cite{Kresse1996,Kresse1996'} using the projector augmented wave (PAW) method\cite{Blochl1994,Kresse1999}. The input structure is taken from experimental results\cite{Mar1992}. We choose the generalized-gradient approximation of Perdew-Burke-Ernzerhof type exchange-correlations\cite{Perdew1996}. An energy cut-off of 360 eV for truncation of the plane wave basis and a 13$\times$9$\times$4 k-mesh is used for the bulk calculation. A Gaussian smearing method was used with a smearing parameter of 50 meV. The spin-orbit coupling is also included in the calculations. The parameters for the tight-binding Hamiltonian are determined from the maximally localized Wannier function (MLWF) method\cite{Mostofi2008,Marzari2012}. We choose the d orbital of tungsten and p orbital of tellurium as the projection. The momentum resolved and spin-projected surface density of states for a semi-infinite slab are then computed using an iterative Green¡¯s function method\cite{Sancho1985}.

\begin{figure}[htb]
\includegraphics[width=8.5 cm]{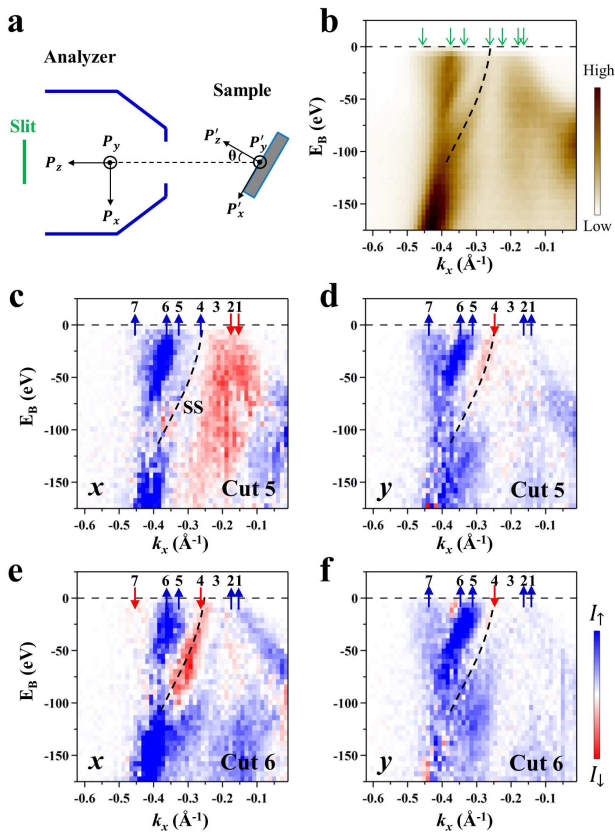}
\caption{(a) Schematic drawing of the SARPES measurement configuration. (b) SARPES intensity plots $I_{tot}$ along cut 5 in the laboratory frame for $S_z$, where $I_{tot}=(I_{\uparrow}+I_{\downarrow})$. The green arrows indicate the bands that cross the Fermi level. (c,d) and (e,f) Spin-resolved ARPES intensity plots $I_{dif}$ in the laboratory frame for S$_x$ and S$_y$ along cut 5 and cut 6, respectively. The blue and red arrows indicate the spin-up and spin-down states for each band at the Fermi level. The angle $\theta$ is 15$^{\circ}$ for (b)-(f).}
\label{fig3}
\end{figure}

WTe$_2$ is a non-magnetic material and has a distorted transition metal dichalcogenide structure. Its space group is classified to Pnm21, as shown in Fig.1(a). Fig.1(b) shows the atomic structure of the natural cleavage plane with the primitive unit cell indicated by the black rectangle. The crystal structure of WTe$_2$ possesses one mirror plane $x$=0 and one glide mirror plane parallel to $y$=0, which transform to k$_x$=0 and k$_y$=0 plane respectively in the momentum space. So the spin texture should respect time reversal symmetry, glide reflection symmetry and mirror symmetry simultaneously.

First, we perform spin-integrated ARPES to capture the whole band structures of WTe$_2$. The Fermi surface consists of several Fermi pockets in the first BZ, as shown in Fig.2(a). Compared with the band structure measured along the $\Gamma$-X axis (Figs.2(b) and S1(a)), we find several bands that cross the Fermi level, as indicated by $k_1$ to $k_7$. Although the intensity of $k_7$ is very weak in the spin-integrated ARPES data in Fig.2, it is clearly observed in the spin-resolved ARPES measurements, as shown in Fig.3. The band structures of the Fermi pockets measured along cuts 2-4 are shown in Figs.2(c)-2(e), and the corresponding second derivative images are shown in Fig.S1. Near the $\Gamma$ point, there are hole-like bands which do not cross the Fermi level (Figs.2(c) and S1(b)). On the other hand, Figs.2(d) and 2(e) show the existence of hole pockets and electron pockets centered on the $\Gamma$-X axis. From these measurements, we conclude that there are two hole pockets (Figs.2(a), 2(b) and S1(a)) and two electron pockets (Fig.3(b)) at each side of the $\Gamma$ point, as indicated in Fig.2(f). It should be noted that the two electron pockets ($k_5$ and $k_6$) are not clearly separated along the $\Gamma$-X axis (cut 1) due to the small momentum separation. However, they are clearly visible along cut 5 (or cut 6), as shown in Fig.3(b). Besides the Fermi pockets, there is a Fermi arc surface state, as indicated by the ``SS" in Fig.2(a). The calculated Fermi surface (Fig.2(g)) and band structure (Fig.2(h)) along the $\Gamma$-X axis are in qualitative agreement with the measured band structures. The Fermi pockets as well as the Fermi arcs in WTe$_2$ have also been reported in several ARPES measurements recently\cite{Bruno2016,WangC2016,WuY2016}.

Next, we performed extensive SARPES to determine the spin texture of WTe$_2$. The definition of the spin directions in the measurements is schematically shown in Fig.3(a). The data directly measured use the laboratory frame ($P_x$, $P_y$, $P_z$) which is related to the sample frame ($P_x^{\prime}$, $P_y^{\prime}$, $P_z^{\prime}$) by the following relationships that can be derived from Euler's rotation theorem:
\begin{eqnarray}\label{equ1}
P_{x}^{\prime}&=& P_{x}cos\theta + P_z sin\theta, \\
P_{y}^{\prime}&=& P_{y}, \\
P_{z}^{\prime}&=& - P_{x}sin\theta + P_z cos\theta ,
\end{eqnarray}
where $\theta$ is the angle between the surface normal and the axis of the analyzer and $P$ is the value of spin polarization. We measured two parallel cuts in the opposite sides of the $\Gamma$-X axis, \textit{i.e.}, cut 5 and cut 6 in Fig.2(a). Fig.3(b) shows the band structure measured along cut 5 in the SARPES mode by summing the intensities of spin-up and spin-down electrons: $I_{tot}=I_{\uparrow}+I_{\downarrow}$. The Fermi pockets and Fermi arcs are clearly resolved, which enables the determination of the detailed spin texture. The seven Fermi wave vectors are also observable, as indicated by the green arrows.

The information of spin polarization can be obtained by the intensity difference of spin-up and spin-down electrons: $I_{dif}=I_{\uparrow}-I_{\downarrow}$. The relationship between $I_{dif}$ and the spin polarization $P$ is: $P=\frac{1}{S_{eff}}\frac{I_{dif}}{I_{tot}}$, where ${S_{eff}}$ is the effective Sherman function which was determined to be 0.25 and 0.3 for the two VLEED detectors from the reference sample. Since the values of ${S_{eff}}$ and $I_{tot}$ are always positive, the spin maps which show the energy and momentum distribution of $I_{dif}$ directly reflect the spin polarization of the bands. Figs.3(c)-3(f) and Fig.S2 show the spin maps in the x, y and z directions along cut 5 and cut 6, respectively, which clearly show the spin polarization of the Fermi arcs and Fermi pockets. The direction of the spin polarization near the Fermi level is indicated by the blue and red arrows for each band.

\begin{figure}[!b]
\includegraphics[width=8 cm]{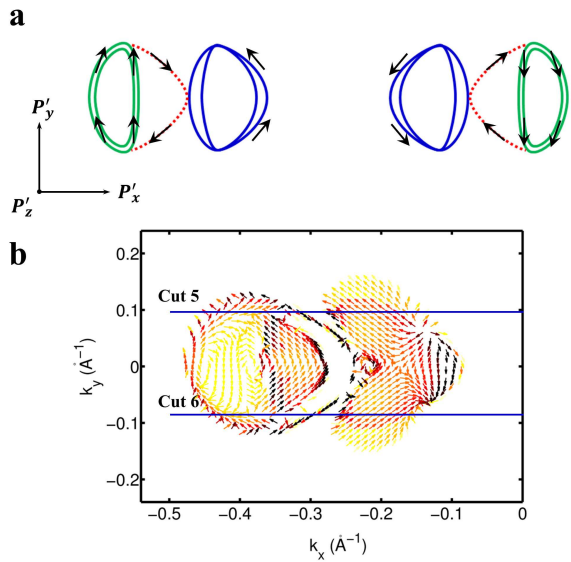}
\caption{(a) Schematic drawing of the measured spin texture of the Fermi surface in the sample frame. (b) Calculated spin texture of the electronic states at the Fermi surface. The arrows show the horizontal spin directions, and the color scheme follows the momentum-resolved surface density of states in Fig.2(g). The blue lines indicate cut 5 and cut 6 that correspond to the measurement positions. The spin polarization of the Fermi arcs and Fermi pockets is clearly found.}
\label{fig4}
\end{figure}

To further understand the spin texture of WTe$_2$, we transform the spin polarization $P$ in the laboratory frame to the sample frame (Fig.3(a)) using equation (1)-(3). In our measurements (Figs.3(b)-3(f)), $\theta$ is 15$^\circ$. The spin polarization maps in the sample frame are shown in Fig.S3. In Fig.4(a), we summarize the spin texture at the Fermi level. The spin texture of the right half of the Fermi surface is calculated based on the crystal symmetry. We find that both the Fermi arc surface states and bulk Fermi pockets are spin polarized. At (k$_x$, k$_y$) and (k$_x$, -k$_y$), the directions of x-spin are opposite while the directions of y-spin are the same, which fulfill the requirements by time reversal symmetry, glide reflection symmetry and mirror symmetry. The calculated spin texture of the Fermi surface (Fig.4(b)) agree well with our experimental results, thus giving strong evidence for the type-\uppercase\expandafter{\romannumeral2} Weyl semimetal state in WTe$_2$.

Based on our experimental and theoretical results, we briefly discuss the mechanism of XMR in WTe$_2$. (\romannumeral1) As a type-\uppercase\expandafter{\romannumeral2} Weyl semimetal, WTe$_2$ possesses pairs of three dimensional Weyl cones. As a result, XMR is expected owing to the linear band dispersion near the Weyl points\cite{Abrikosov1998} or the interplay of impurities with the Weyl fermions\cite{Ramakrishnan2015}. (\romannumeral2) Based on the spin texture we measured (Fig.4), if quasiparticles are scattered by non-magnetic impurities, a lot of scattering channels are prohibited due to the small overlapping of the initial states and final state, leading to a small intrinsic resistivity. The prohibited scattering channels could open in the presence of magnetic fields\cite{Jiang2015}, generating the XMR in WTe$_2$. So our results show that both mechanisms might contribute to the large and non-saturating MR in WTe$_2$.

Our combined experimental and theoretical results support the existence of topological non-trivial Fermi arcs in WTe$_2$, in agreement with several recent works\cite{XuN2016,WangC2016,WuY2016,WangZ2016}. However, recently Tamai {\it et al.} reported that the topological properties of the Fermi arcs in WTe$_2$ and MoTe$_2$ strongly depend on the spin-orbit coupling (SOC) strength\cite{Tamai2016}. By artificially modifying the SOC strength, the spin-polarized Fermi arcs can be either topological trivial or nontrivial\cite{Tamai2016} while keeping a similar shape, making the validation of the topological Fermi arcs a challenging task. As a result, further efforts are highly demanded to clarify this debate.

In summary, we investigated the band structure and spin texture of WTe$_2$ by SARPES and first-principles calculations. Our results reveal the existence of spin-polarized Fermi arcs and thus support WTe$_2$ as a type-\uppercase\expandafter{\romannumeral2} Weyl semimetal candidate. Moveover, we also observe the spin polarization of the Fermi pockets, which respect the TRS and mirror symmetry. These results also provide important information to understand the large and non-saturating MR in WTe$_2$.

\begin{acknowledgements}
We thank Tay-Rong Chang for the discussion of the MLWF method and thank C. Ibrahim for the help in the experiments. We also thank Prof. X. J. Zhou for providing the igor macro to analyze the ARPES data. This work was supported by the Ministry of Education, Culture, Sports, Science and Technology of Japan (Photon and Quantum Basic Research Coordinated Development Program), a Japan Society for the Promotion of Science grant-in-aid for specially promoting research (\#23000008) and for Scientific Research (B) (\#26287061), by the U.S. Department of Energy (DOE), Office of Science (OS), Office of Basic Energy Sciences, Division of Materials Science and Engineering, under Grant No. DE-FG02-07ER46383 (T.-C.C.), by the U.S. National Science Foundation Grant No. 1542747, and by the Thematic Project at the Academia Sinica, by the Strategic Priority Research Program (B) of the Chinese Academy of Sciences (No.XDB07020100), and the National Natural Science Foundation of China (No.11474330,11274367).
\end{acknowledgements}

\end{document}